\documentclass{article}

\usepackage[english]{babel}

\usepackage[letterpaper,top=3cm,bottom=3cm,left=3cm,right=3cm,marginparwidth=1.75cm]{geometry}

\usepackage{graphicx}

\usepackage[hidelinks, colorlinks=true, citecolor=blue, linkcolor=blue]{hyperref}

\usepackage{microtype}
\usepackage{mathtools}

\usepackage{amsthm}
\usepackage{amssymb}
\usepackage{mathabx}
\usepackage{authblk}
\usepackage{dirtytalk}
\usepackage{algorithmic}
\usepackage{algorithm}
\usepackage{booktabs}
\usepackage{stmaryrd}
\usepackage[shortlabels]{enumitem}
\usepackage{listings}
\usepackage{thmtools} 
\usepackage{thm-restate}

\newtheorem{definition}{Definition}

\newtheorem{corollary}{Corollary}

\theoremstyle{remark}

\newcommand{\mech}{\mathcal{M}}
\newcommand{\complex}{\mathbb{C}}
\newcommand{\real}{\mathbb{R}}
\newcommand{\Exp}{\mathbb{E}}

\DeclareMathOperator{\MSE}{MSE}
\DeclareMathOperator{\Var}{Var}
\newcommand{\D}{\mathcal{D}}
\newcommand{\eg}{e.g.\@ }
\newcommand{\ie}{i.e.\@ }

\usepackage[frozencache,cachedir=.]{minted}

\usepackage{csquotes}

\begin{document}

\title{\textbf{Complex-valued Federated Learning with Differential Privacy and MRI Applications}}

\author[1,2]{Anneliese Riess}
\author[2]{Alexander Ziller}
\author[3]{Stefan Kolek}
\author[2,4]{Daniel Rueckert}
\author[1,2,5]{Julia Schnabel}
\author[1,2,4]{Georgios Kaissis}

{\small
\affil[1]{Helmholtz Munich, Germany}
\affil[2]{TUM, Germany}
\affil[3]{LMU Munich, Germany}
\affil[4]{Imperial College London, UK}
\affil[5]{King’s College London, UK}
}

\date{}
\maketitle    

\begin{abstract}
Federated learning enhanced with Differential Privacy (DP) is a powerful privacy-preserving strategy to protect individuals sharing their sensitive data for processing in fields such as medicine and healthcare. 
Many medical applications, for example magnetic resonance imaging (MRI), rely on complex-valued signal processing techniques for data acquisition and analysis.
However, the appropriate application of DP to complex-valued data is still 
underexplored.
To address this issue, from the theoretical side, we introduce the complex-valued Gaussian mechanism, whose behaviour we characterise in terms of $f$-DP, $(\varepsilon, \delta)$-DP and Rényi-DP.
Moreover, we generalise the fundamental algorithm DP stochastic gradient descent to complex-valued neural networks and present novel complex-valued neural network primitives compatible with DP.
Experimentally, we showcase a proof-of-concept by training federated complex-valued neural networks with DP on a real-world task (MRI pulse sequence classification in $k$-space), yielding excellent utility and privacy.
Our results highlight the relevance of combining federated learning with robust privacy-preserving techniques in the MRI context. 
\bigbreak
\end{abstract}

\section{Introduction}
Complex-valued (CV) signal processing is becoming increasingly important in medicine and medical imaging tasks.
For example, frequency-domain magnetic resonance imaging (MRI) data is acquired in the CV $k$-space; audio signals from speech or a patient’s heartbeat can be represented as CV spectrograms and wearable or implanted biological sensors produce measurements which can be efficiently represented in the complex field.
Moreover, many real-valued problems can be represented and solved in the complex domain, \eg differential equations, which can be solved more efficiently by first taking their Fourier/Laplace transforms.
In many of these examples, privacy preservation is paramount to protect patients and to furnish objective security guarantees, and is often mandated by ethical considerations and legal regulations \cite{jobin2019global}.

Federated learning (FL) has been proposed to enable privacy-preserving data processing in medical imaging.
Here, users contribute to training a joint model without sharing their private data, but rather only model updates (\eg gradients) with a central server that coordinates the training.
However, this decentralized approach alone does not suffice to prevent privacy violations, as prior works have shown that FL models are vulnerable to attacks which disclose sensitive information, such as data reconstruction attacks \cite{boenisch2023curious,feng2024privacy,fowl2022robbing}.
At the same time, FL does not provide a formal privacy guarantee that can objectively quantify the protection provided by this approach.
These remarks underscore that, to be formally privacy-preserving, FL must be complemented by additional privacy technologies. 
Differential Privacy (DP) \cite{dwork2014algorithmic}, a formal framework and set of techniques for deriving insights from sensitive databases while protecting the privacy of individuals who contributed their data, has established itself as the tool of choice in this regard. 
DP can be thought of as a \say{contract} between a data owner and a data processor that guarantees that the influence of any individual’s data on the outcome of a computation and --by extension-- any harm originating from the release of its results, is limited.
However, while DP has been studied extensively in many sub-fields of AI, to our knowledge, it has yet not been sufficiently investigated in the context of CV data processing.

\paragraph{Our Contributions}
We propose key theoretical and methodological innovations to enable the application of DP in federated CV neural networks (CVNNs). 
Concretely, we (1) introduce the complex-valued Gaussian mechanism and characterise its privacy properties in Section \ref{sec::theoreticalresults}; (2) we generalise the fundamental algorithm of DP deep learning, DP stochastic gradient descent (DP-SGD) to CVNNs in Section \ref{sec::complexdpsgd}; (3) we propose novel CVNN primitives (complex GroupNorm and ConjMish, a new activation function) and investigate their properties in Section \ref{sec::cvnnprimitives}; (4) finally, in a proof-of-concept medical imaging CV FL application in Section \ref{sec:experiments}, we find that that applying our methods yields excellent accuracy.

\paragraph{Related Work}
CV signal processing workflows have witnessed increasing interest over the past few years. 
Arguably, biosensing \cite{peker2016efficient} and magnetic resonance imaging analysis \cite{cole2020analysis,kustner2020cinenet,virtue2017better} are among the most relevant for privacy preservation, and have also seen increasing usage of AI tools.
CVNNs have only recently gained significant traction, as automatic differentiation systems have --until recently-- not natively supported CV gradients and due to the increased computational expense of CV operations. 
This has changed with the near-universal adoption of the Wirtinger calculus \cite{kreutz2009complex,Wirtinger1927} in deep learning frameworks, and with the introduction of native (i.e. hardware-optimised) primitives for \eg convolutions.
So far, only a single other study has demonstrated the use of CVNNs in FL \cite{yu2024complex}, and no studies before ours have addressed the biomedical domain or DP applications therein.

DP \cite{dwork2006calibrating} has become a standard technique for privacy preservation in AI. 
Due to space constraints, for a detailed introduction to DP we refer to \cite{dwork2014algorithmic,dong2019gaussian}.  
DP-SGD for real-valued NNs was introduced by \cite{abadi2016deep}. 
Only a limited number of studies have examined DP in conjunction with CV data \cite{chatalic2022compressive,fan2013adaptively,fioretto2019differential} or introduced techniques for privacy \textit{accounting} using CV functional representations \cite{koskela2020computing}, however, to our knowledge, none have formalised a general framework to handle DP for CV tasks.

\section{Preliminaries}
\label{sec::preliminaries}
Throughout the paper, we assume a standard FL setup with a \textit{central server} and several \textit{computation nodes}, but all introduced techniques apply equally to peer-to-peer FL topologies, swarm learning, etc..
Moreover, we assume all parties to be \textit{honest but curious}, such that computation nodes perform a local privatisation of their updates before submitting them to the central server; this is not a limitation as our techniques can be readily adapted to all other threat models.
As is customary in DP literature, each node holds a set of sensitive records from a universe $\mathcal{X}$, called a database $\D$. 
From this, an adjacent database $\D'$ can be constructed by adding, removing or replacing the data of a single individual. 
We assume without loss of generality that individuals are unique throughout the federation.
DP is typically realised by first executing a deterministic query function $q$ over the database and then randomising its output by the addition of noise through a DP mechanism $\mech$. 
The noise is calibrated to the query function’s (global) ($\ell_p$-)sensitivity induced by the $p$-norm ($p \in [1, \infty)$), which we denote by $\Delta_p (q)$.
As the $p$-norm is also defined for CV vectors, we employ the same strategy to randomise the output of a CV query.
In turn, we introduce the complex Gaussian mechanism (cGM) in Section \ref{sec::theoreticalresults}, which serves as the CV counterpart to the Gaussian mechanism (GM), one of the most employed mechanism to achieve DP in real-valued settings.  

Every CV function $q: \mathcal{X} \to \complex^n$ can be written as: 
\begin{equation}
    q(\D) = q_{\Re}(\D) + \text{i} \cdot  q_{\Im}(\D), \label{eq::representationcomplexreal}
\end{equation}
where $q_{\Re}:\mathcal{X} \to \real^n$ and $q_{\Im}:\mathcal{X} \to \real^n$ denote the real and imaginary parts of $q(\D)$, respectively. 
Representation \ref{eq::representationcomplexreal} is useful as many complex functions can be thought of as operators acting on the real and imaginary parts of a complex number separately and then \say{assembling} the result.
However, we caution against equating CV networks to real-valued networks with two \say{channels} if the appropriate CV operations are not used, as this discards the information within the relationship between the real and imaginary part.
Other differences between $\complex$ and $\real^2$ call for distinctive treatment when handling tasks in $\complex$.
For instance, since the complex plane does not admit a natural ordering, the minimisation of CV functions is not defined. 
Hence, CVNNs use complex-to-real loss functions.
Moreover, to obtain correct gradients for optimisation, we employ the Wirtinger calculus \cite{kreutz2009complex}, which provides a CV gradient operator for real-valued loss functions in CVNNs. 
In particular, this serves as a base for our proposed CV counterpart to DP-SGD.

\section{Theoretical Results}
\label{sec::theoreticalresults}
To characterise the privacy properties of CV mechanisms, we utilise the $f$-DP framework \cite{dong2019gaussian}.
Relying on statistical hypothesis testing, $f$-DP interprets DP through a \textit{trade-off function} $T$ between the \textit{Type I} and \textit{Type II} statistical errors faced by a membership inference adversary trying to determine whether one of the adjacent databases contains the individual or not.  
A mechanism $\mathcal{M}$ satisfies $f$-DP if, for all pairs of adjacent databases $\D$ and $\D'$, $T(q(\D), q(\D'))(\alpha) \geq f(\alpha)$ holds $\forall \alpha \in [0,1]$ for some trade-off function $f$. 
Gaussian DP (GDP) is a specialisation of $f$-DP when the trade-off-function has the form $G_\mu \coloneqq T\left( \mathcal{N}(0,1),\mathcal{N}(\mu,1) \right)$. 
In particular, $\mathcal{M}$ preserves $\mu$-Gaussian DP ($\mu$-GDP) if it preserves $f$-DP, for $f(\alpha) = G_\mu(\alpha) = \Phi \left(\Phi^{-1}(1-\alpha) -\mu) \right)$, where $\alpha$ is the Type-I statistical error and $\Phi$ is the cumulative distribution function of the standard, real-valued normal distribution.
In this light, we introduce our key CV additive noise mechanism:
\begin{definition}[Complex Gaussian mechanism]
    Let $q:\mathcal{X} \to \complex^n$, $\D \in \mathcal{X}$, and $\psi \sim \mathcal{N}_\complex\left( \textbf{0}, \Gamma, C\right)$ denote the complex Gaussian distribution with location parameter $\mu =\textbf{0} \in \mathbb{C}^n$, covariance matrix $\Gamma \in \mathbb{C}^{n\times n}$ and relation matrix $C \in \mathbb{C}^{n\times n}$.
    Then, the complex Gaussian mechanism (cGM) is defined as:
    \begin{equation}
        \mech(\D) = q(\D) + \psi.
    \end{equation}
\end{definition}

We will consider the cGM when $\psi \sim \mathcal{N}_\complex\left( \textbf{0}, 2 \sigma^2  \mathbf{I}_n, 2 i  \gamma  \mathbf{I}_n\right)$.
The cGM has variance $\sigma^2$ in the real and imaginary part of each coordinate and \textit{total} variance $2\sigma^2$ per coordinate, as $\Var(z) = \Var(\Re(z)) + \Var(\Im(z))$ holds for any random variable in $\complex$.
Moreover, the cGM marginals are \textit{non-circular} complex Gaussian distributions whose real and imaginary components are correlated with correlation coefficient $\rho = \frac{\gamma}{\sigma^2}$.
Whenever $\rho = 0$, we observe a special case: the \textit{circular} cGM, whose marginals are circular complex Gaussian distributions. 
In turn, its real and imaginary components are i.i.d.\@ scalar real-valued Gaussian distributions $\mathcal{N}(0, \sigma^2)$.
We next characterize the privacy properties of the cGM:

\begin{restatable}{theorem}{tradeoffgausscomplex} \label{thm::gdpgausscomplexdependent}
    Let $\mech$ be the cGM with correlation coefficient $\rho \neq 1$ acting on a query function $q$.
    Then, $\mech$ satisfies $\mu$-GDP with:
    \begin{equation}
        \mu = \sqrt{\frac{\Delta_2(q)^2}{\sigma^2(1-\rho^2)} + \frac{2 |\rho|}{\sigma^2 (1-\rho^2)} \cdot \Delta_2(q_\Re) \cdot \Delta_2(q_\Im)}.
        \label{eq::newmudependent}
    \end{equation}
\end{restatable}
\begin{proof}
    The proof of Theorem \ref{thm::gdpgausscomplexdependent} can be found in Appendix \ref{apx::proof1}.
\end{proof}
Note that \ref{eq::newmudependent} is monotonically increasing in $|\rho|$. 
Specifically, for $|\rho|\to 1$, $\mu \to \infty$ and the cGM becomes \textit{blatantly non-private}
We next turn to the circular case:
\begin{corollary} \label{thm::tradeoffgausscomplex}
    The circular cGM acting on $q$ satisfies $\mu$-GDP with $\mu = \Delta_2(q)/ \sigma$.
\end{corollary}
\begin{proof}
    Follows from Theorem \ref{thm::gdpgausscomplexdependent} by setting $\rho=0$.
\end{proof}
Since the cGM (including the circular special case) satisfies $\mu$-GDP, it inherits all of the properties of GDP, \ie resilience to post-processing, group privacy, subsampling and composition.
Additionally, one can provide $(\varepsilon,\delta)$ and R\'{e}nyi-DP guarantees using the techniques presented in \cite{dong2019gaussian,Mironov_2017}.
These results allow one to leverage available privacy accounting tools for real-valued DP to design DP workflows for CV tasks as demonstrated in Section \ref{sec:experiments}.

Interestingly, Corollary \ref{thm::tradeoffgausscomplex} shows that choosing $\rho \neq 0$ can \textit{never improve the privacy guarantee of the cGM}.
Moreover, observe that the mean squared error (MSE) between $z \in \complex$ and its perturbed version $z + \psi$, where $\psi$ is a CV random variable drawn from a zero-centered distribution, satisfies: 
\begin{align}
    \MSE \left(z,z + \psi\right) = \Exp\left(||z-(z + \psi)||_2^2\right) = \Var(\psi) = \Var ( \Re(\psi)) +  \Var ( \Im(\psi)). \label{eq::formMSE}
\end{align} 
Hence, the MSE is independent of the correlation between the real and imaginary components of the CV noise, and consequently, there is no benefit from using correlated complex noise in terms of the introduced distortion.
The circular cGM is thus in this sense \textit{optimal} in terms of its privacy-utility trade-off.

\section{Training CVNNs with DP} 
\label{sec:neuralnets}
\paragraph{$\zeta$-DP-SGD} \label{sec::complexdpsgd}
Real valued DP-SGD \cite{abadi2016deep} is a key technique to train deep NNs with DP.
Recall that the key steps of DP-SGD are (1) clipping the $\ell_2$-norm of the \textit{per-sample} gradients to a pre-defined threshold, and (2) adding (real-valued) Gaussian noise calibrated to this threshold.
Then, each training step leads to the release of a privatised gradient which is used to update the local node's weights, or is shared with the central server, \eg for aggregation. 
To generalize DP-SGD to CVNNs and enable their federated training, we next introduce $\zeta$-DP-SGD, presented in Algorithm \ref{dp-sgd}.
Recall from Section \ref{sec::preliminaries} that a complex-to-real loss function $\mathcal{L}: \mathbb{C}^n \mapsto \mathbb{R}^1$ is minimised in CVNNs.
Using the Wirtinger calculus, we clip the $\ell_2$-norm of the per-sample \textit{conjugate gradient} \cite{kreutz2009complex}, which represents the direction of steepest ascent in this setting:
\begin{equation}
    \nabla \overline{\mathcal{L}} \coloneqq 2 \left (\frac{\partial \mathcal{L}}{\partial \overline{\theta}_1}, \dots, \frac{\partial \mathcal{L}}{\partial \overline{\theta}_n} \right),
\end{equation}
where $\boldsymbol{\overline{\theta}} = \left(\overline{\theta}_1, \dots, \overline{\theta}_n\right)$ is the conjugate weight vector. 
The conjugate gradient is twice the conjugate Wirtinger derivative with respect to the weights \cite{kreutz2009complex}, which results in parity with the real-valued case in terms of the effective learning rate.
In particular, using the theoretical results in the previous section, each step of $\zeta$-DP-SGD satisfies GDP, enabling us to utilise the composition and sub-sampling theorems of \cite{dong2019gaussian} to account for the total privacy cost of training a CVNN.

\begin{algorithm}[ht]
	\caption{$\zeta$-DP-SGD}
	\begin{algorithmic}
	\REQUIRE Database with samples $\{z_1,\ldots,z_N\} \in \mathbb{C}^n$, neural network with loss function $\mathcal{L}$ and weight vector $\boldsymbol{\theta} \in \mathbb{C}^m$. Hyperparameters: learning rate $\eta_t$, noise variance $\sigma^2$, sampling probability $p=\frac{R}{N}$, gradient norm bound $B$, total steps $T$. 
		\STATE {\bf Initialize} $\boldsymbol{\theta}_0$ randomly
		\FOR{$t \in [T]$}
		\STATE {Draw a batch $L_t$ with sampling probability $p$ (\eg using \textit{Poisson} sampling)}
		\STATE {\bf Compute per-sample conjugate gradient}
		\STATE {For each $i\in L_t$, compute $\overline{\boldsymbol{g}}_t(z_i) \gets \nabla \overline{\mathcal{L}}(\boldsymbol{\theta}_t, z_i)$}		
		\STATE {\bf Clip conjugate gradient}
		\STATE {$\widecheck{\boldsymbol{g}}_t(z_i) \gets \overline{\boldsymbol{g}}_t(z_i) / \max\left(1, \frac{\Vert \overline{\boldsymbol{g}}_t(z_i)\Vert_2}{B}\right)$}
		\STATE {\bf Apply the circular complex Gaussian Mechanism and average}
		\STATE {$\widetilde{\boldsymbol{g}}_t \gets \frac{1}{R}\left( \sum_i \widecheck{\boldsymbol{g}}_t(z_i) + \mathcal{N}_{\mathbb{C}}(\boldsymbol{0}, 2B^2\sigma^2 \mathbf{I}_m, \boldsymbol{0})\right)$}
		\STATE {\bf Descend}
		\STATE { $\boldsymbol{\theta}_{t+1} \gets \boldsymbol{\theta}_{t} - \eta_t \widetilde{\boldsymbol{g}}_t$}
		\ENDFOR
		\STATE {\bf Output} updated neural network weight vector $\boldsymbol{\theta}_T$ and compute the privacy cost.
	\end{algorithmic}
	\label{dp-sgd}
\end{algorithm}

\paragraph{CVNN Primitives for $\zeta$-DP-SGD Training} \label{sec::cvnnprimitives}

Many CVNN components such as complex convolutional and linear layers as well as split (\eg $\mathbb{C}$ReLU \cite{trabelsi2017deep}) and fully complex (\eg Cardioid \cite{virtue2017better}) activation functions are compatible with $\zeta$-DP-SGD. 
However, Batch Normalisation (BN) \cite{ioffe2015batch} and its CV implementation \cite{trabelsi2017deep} are prohibited in DP as they \say{contaminate} the activations with information from other samples in the batch, leading to undefined \textit{per-sample} (conjugate) gradients, which are required for a correct implementation.

To address this issue, BN is typically replaced with Group Normalisation (GN) \cite{wu2018group} in DP NNs. 
Since a CV implementation of the GN layer is missing, we next introduce a novel \textbf{CV GN layer}.
Recall that, while real vectors are normalised by subtracting the mean and dividing by the variance, in complex vectors the covariance between the real and imaginary components must also be considered.
We address this by grouping the activations, and then \textit{whitening} them group-wise.
Similar to \cite{trabelsi2017deep}, we initialise the affine parameters of the GN layer to $\gamma=\frac{1}{\sqrt{2}} + \frac{1}{\sqrt{2}}\text{i}$ and to $\beta=\boldsymbol{0}$. 
An implementation of the whitening algorithm and of the GN layer can be found in Listing \ref{ls:whitening}.
Of note, the same approach can be used for Layer, Instance or weight normalisation \cite{salimans2016weight}, as our implementation is differentiable. 

\begin{listing}[ht] 
\caption{PyTorch implementations of complex GN.}
\begin{minted}[breaklines=True, fontsize=\scriptsize]{Python}
def whiten_single(vec):
    flat_vector = vec.flatten()
    centered = flat_vector - flat_vector.mean() # subtract mean to center the tensor
    stacked = torch.stack([centered.real, centered.imag])
    sigma = torch.cov(stacked) # compute covariance between real and imaginary.
    u_mat, lmbda, _ = torch.linalg.svd(sigma) # Compute 1/sqrt. of covariance matrix.
    w_mat = torch.matmul(
        u_mat, torch.matmul(torch.diag(1.0 / torch.sqrt(lmbda + 1e-5)), u_mat.T))
    result = torch.matmul(w_mat, stacked)
    return (result[0] + result[1] * 1j).reshape(vec.shape)

whiten_group = vmap(vmap(whiten_single)) #vmap over batch and group axis

class ComplexGroupNorm2d(nn.Module):
    def __init__(self, num_groups, num_channels):
        ... #initialise gamma and beta

    def forward(self, x):
        group_shape = (-1, self.groups, self.num_channels // self.groups) + x.shape[2:]
        x = x.reshape(group_shape) # split into groups
        x = whiten_group(x) # whiten each group
        x = x.reshape((-1, self.num_channels,) + group_shape[3:]) # reshape to original shape
        x = x * gamma + beta # affine operation
        return x
\end{minted}
\label{ls:whitening}
\end{listing}
As an additional contribution, we introduce a novel \textbf{CV Mish activation function}. 
Recall that the real-valued Mish \cite{misra2019mish} is defined as: $\operatorname{Mish}(x) \coloneqq  x \tanh{\left(\log{\left(e^{x} + 1 \right)}/x \right)}$.
For use with CVNNs, we define a \textit{conjugate} version:
\begin{equation}
    \operatorname{ConjMish}(z) \coloneqq \left(1 + \text{i}\right) \operatorname{Mish}(\Re(z)) -
    \left(1 - \text{i}\right) \operatorname{Mish}(\Im(z)).
\end{equation}
\noindent
We empirically found ConjMish to drastically improve accuracy by up to 5\% over the best previous alternatives (Cardioid \cite{virtue2017better}, ModReLU or $\mathbb{C}$ReLU \cite{trabelsi2017deep}). 
In contrast to Cardioid and $\mathbb{C}$ReLU, ConjMish has both a \textit{magnitude thresholding} effect and  \say{phase non-linearity} effect instead of merely \say{passing through} the phase.
The latter seems to improve NN convergence, and could be of independent interest. 
We leave a detailed investigation to future work.

\section{Experiments} \label{sec:experiments}

We next demonstrate the experimental evaluation of our framework in the context of training federated CVNNs on a real-life medical dataset, where both, stringent privacy guarantees and high accuracy are desired. 
We selected the task of automated MRI pulse sequence classification, which is relevant for both, the automated curation of medical images for AI applications and for image retrieval tasks in clinical routine.
Recent works have tackled this challenge using both supervised \cite{VieiradeMello2021} and unsupervised \cite{Kart2021} deep learning techniques. 
Contrary to the aforementioned works, we directly classify the MRI pulse sequence in $k$-space, that is, to \textit{directly} classify the CV frequency-domain MRI data. 

We utilised data from the \textit{brain} sub-challenge of the Medical Segmentation Decathlon \cite{antonelli2021medical}, consisting of $484$ training records and $266$ test records, which are partitioned such that one patient is only present in a single dataset.
We instantiated an FL simulation using the \texttt{Flower} framework \cite{beutel2020flower} which we augmented with a customised version of \texttt{Opacus} \cite{opacus}, and distributed the training records uniformly at random among $11$ computation nodes to obtain an i.i.d.\@ FL setting.
Moreover, we uniformly distributed $110$ randomly selected test records to each computation node to serve as a validation set.
The rest of the test set remained at the central server and was used only to compute the final accuracy.
Additionally, for comparison, we also trained the CVNN under centralised conditions.
From each record, we extracted $20$ centre slices for each of the four available pulse sequences: Fluid Attenuation Inversion Recovery (FLAIR), T1-weighted (T1w), T1-weighted with contrast agent (T1wGD) and T2-weighted (T2w).
This resulted in a total dataset size of $38\,720$ training and $21\,280$ testing images, which we resized to $32 \times 32$ pixels, Fast Fourier transformed to simulate $k$-space (where we retained duplicated frequency components to obtain a representation with the same dimensions as the input), and normalised by whitening.

We used the model architecture from \cite{dormann2021not} consisting of three convolutional blocks with $32$, $64$ and $128$ filters, CV GN and the ConjMish activation function (see Section \ref{sec::cvnnprimitives}) as well as average pooling layers.
The classification layer of the network consisted of a single linear layer with $128$ units. 
We trained the model using Adaptive Federated Averaging \cite{reddi2021adaptivefederatedoptimization} for $500$ epochs using the NAdam optimiser with a learning rate of $0.0002$, which we decayed by $10\times$ after $300$ epochs, a fixed $\ell_2$-norm bound of $1$, $16$ GN groups, an expected batch size of $24$, and a target $\varepsilon \in [1, 3, 5, 10]$ for $\delta=0.001$.
All stated privacy guarantees are \say{per-patient}, and all nodes participated in every round with one round per step.
Table \ref{tab::mri} summarises these results across $10$ random seeds as well as the results from centralised learning as a reference.
We note for completeness that the $\varepsilon$-parameter represents the \textit{privacy budget} in DP, and higher values correspond to worse privacy guarantees for the individuals.
Interestingly, the CVNN achieved an accuracy of nearly $90\%$ at an $\varepsilon$-value of $3$, with (at most) $1$ to $2\%$ of additional performance gained by diminishing the privacy guarantee.
This indicates that, in the task we consider, relatively stringent (local) DP guarantees can be achieved in FL practically without any accuracy penalty, even compared to centralised learning.

\begin{table}[h] 
\caption{Accuracy in $\%$ (mean $\pm$ standard deviation) on the MRI test set across $10$ random seeds for FL and centralised learning (CL).}
\centering
\begin{tabular}{@{ }l  llllll@{ }} 
\toprule
$\varepsilon$ & $1$      & $3$      & $5$       & $8$       & $10$    & $\infty$   \\ \midrule
FL         & $81.65 \pm 1.03$\;  & $87.98 \pm 1.46$\;  & $88.46 \pm 1.37$\;   & $89.08 \pm 1.11$\;   & $89.99 \pm 0.70$\; & $90.12 \pm 1.26$    \\ 
CL\;          & $82.85 \pm1.49$   & $89.53 \pm 1.89$   & $89.31 \pm 1.49$    & $89.62 \pm 1.06$    & $90.33 \pm 0.59$  & $90.89 \pm 1.41$     \\ \bottomrule    
\end{tabular}
\label{tab::mri}
\end{table}

\section{Discussion}
In this work, we investigated the application of DP techniques to CVNNs.
We theoretically showed that the cGM naturally extends its real-valued counterparts to the complex domain, allowing for efficient privacy accounting. 
Moreover, we experimentally demonstrated a proof-of-concept for FL with DP in CVNNs and found that DP CVNN training is possible with strong privacy guarantees and excellent utility, a crucial combination in sensitive fields like healthcare.

We foresee several interesting avenues for future work:
For one, the communication efficiency of CVNNs in FL is reduced to their nominally higher number of parameters (two real-valued floating point numbers per parameter).
Thus, optimising e.g.\@ mixed-precision techniques for such applications could reduce communication overhead.
Moreover, network quantisation strategies tailored to CVNNs could further increase efficiency while maintaining high accuracy. 
Furthermore, we intend to explore additional CV mechanisms, such as the CV Laplace mechanism, in future studies.

In conclusion, we anticipate the adoption of CVNNs to increase in a variety of machine learning tasks through the broader availability of software tools and improved hardware support.
In particular, since, in collaborative and federated learning, many such tasks concern sensitive data, we contend that integrating rigorous privacy techniques such as DP is essential for increasing trust by providing formal guarantees of model behaviour.

\bibliographystyle{alpha}

\bibliography{main}

\appendix
\section{Proof of Theorem \ref{thm::gdpgausscomplexdependent}.} \label{apx::proof1}
\tradeoffgausscomplex*
\begin{proof}
Consider a membership inference adversary who observes a mechanism output $y = \mathcal{M}(q(\cdot)) \in \mathbb{C}^n$ and wants to assess whether $y$ originated under $\D$ or $\D'$ based on this single observation.
Moreover, assume the adversary is able to conduct a Neyman-Pearson optimal hypothesis test to distinguish $\mathcal{M}(q(\D))$ from $\mathcal{M}(q(\D'))$.
The proof is thus reduced to a CV simple vs.\@ simple binary hypothesis testing problem for the location parameter (\ie mean) of a complex Gaussian distribution with equal covariance and relation matrix.
Choosing the likelihood ratio as our test statistic leads to the optimal test design with the hypotheses:
    \begin{align}
        \mathcal{H}_0: y \sim \mathcal{N}_{\mathbb{C}}(q(\D), 2 \sigma^2 \mathbf{I}_n, 2\text{i} \gamma \mathbf{I}_n) 
        \quad \text{and} \quad
        \mathcal{H}_1: y \sim \mathcal{N}_{\mathbb{C}}(q(\D'), 2 \sigma^2 \mathbf{I}_n, 2 \text{i} \gamma  \mathbf{I}_n).
    \end{align}  
Now, we introduce some notation to ease reading of the remaining of the proof.
Let $\tilde{z}$ denote the augmented vector constructed from $z \in \complex^n$ in the following way:
\begin{equation}
    \widetilde{z} =         
    \begin{bmatrix}
        z\\
        \bar{z}
    \end{bmatrix} \in \complex^{2n},
\end{equation}
where $\bar{z}$ is the element-wise complex conjugate of $z$.
Moreover, let $C, \Gamma \in \complex^{n\times n}$ be the diagonal matrices  $\Gamma = 2\sigma^2 I_n$ and $C= 2 \text{i} \cdot \gamma I_n$.
Then, the probability density functions (PDFs) $f_0(z)$ and $f_1(z)$ under $\mathcal{H}_0$ and $\mathcal{H}_1$, respectively, are:
    \begin{align*}
        &f_0(z) = \frac{1}{\pi^n\sqrt{\det(\Gamma) \cdot \det(P))}} \exp\left( -\frac{1}{2}[\widetilde{z}-\widetilde{q(D)}]^H
        \begin{bmatrix}
            \Gamma &C\\
            \bar{C} &\bar{\Gamma} \\
        \end{bmatrix}^{-1}
        [\widetilde{z}-\widetilde{q(D)}]
        \right), \\
        &f_1(x) = \frac{1}{\pi^n\sqrt{\det(\Gamma) \cdot \det(P))}} \exp\left( -\frac{1}{2}[\widetilde{z}-\widetilde{q(D')}]^H
        \begin{bmatrix}
            \Gamma &C\\
            \bar{C} &\bar{\Gamma} \\
        \end{bmatrix}^{-1}
        [\widetilde{z}-\widetilde{q(D')}]
        \right),
    \end{align*}
where $P= \bar{\Gamma} - C^H\Gamma^{-1}C \in \complex^{n\times n}$, and $H$ denotes the complex conjugate transpose.
These probability density functions (PDFs) $f_0(z)$ and $f_1(z)$ under $\mathcal{H}_0$ and $\mathcal{H}_1$, respectively, are used to compute the log likelihood ratio $ L = \log\left( \frac{f_1(x)}{f_0(x)} \right)$.
It is well-known that, for this problem, the log-likelihood ratio test statistic (i.e.\@ privacy loss random variable) $L$ is (real) Gaussian distributed \cite{schreier2010statistical}. 
In particular, the mean of $L$ is $\frac{d}{2}$ under $\mathcal{H}_1$ and $-\frac{d}{2}$ under $\mathcal{H}_0$, and its variance is $d$ under both hypotheses, where $d$ is given by:
    \begin{equation}
        d= [\widetilde{q(D')}-\widetilde{q(D)}]^H 
        \begin{bmatrix}
            \Gamma &C\\
            \bar{C} &\bar{\Gamma} \\
        \end{bmatrix}^{-1} 
        [\widetilde{q(D')}-\widetilde{q(D)}], \label{eq::definitiondeflectionspecialcase}    
    \end{equation}
(we refer to Section 7 from \cite{schreier2010statistical} for more details).
Moreover, it is also well-known that that the power of any such a test is monotonically increasing with $d$.
Therefore, we seek to maximise $d$ for all $\D, \D' \in \mathcal{X}$.  
Due to the specific form of the matrix above (see Equation \ref{eq::definitiondeflectionspecialcase}), after some algebraic manipulation, we can compute it explicitly:
    \begin{align}
        \begin{bmatrix}
            \Gamma &C\\
            \bar{C} &\Gamma\\
        \end{bmatrix}^{-1} 
        = 
        \begin{bmatrix}
           \frac{\sigma^2}{2( \sigma^4 - \gamma^2)}\mathbf{I}_n &-\frac{\text{i}\gamma}{2 (\sigma^4 - \gamma^2)}\mathbf{I}_n\\
            \frac{\text{i}\gamma}{2 (\sigma^4 - \gamma^2)}\mathbf{I}_n &\frac{\sigma^2}{2( \sigma^4 - \gamma^2)} \mathbf{I}_n\\
        \end{bmatrix}. \label{eq::inverseaugmentedcovariancematrix}
    \end{align}  
Using \ref{eq::inverseaugmentedcovariancematrix}, we can rewrite $d$, and obtain:
    \begin{equation}
    \begin{split}
        d
        = \frac{||q(\D)-q(\D')||_2^2}{\sigma^2(1-\rho^2)}   
        \pm \frac{2|\rho|}{\sigma^2(1-\rho^2)}(\Re(q(\D)-q(\D')))^T(\Im(q(\D)-q(\D')))
        \label{eq::dsimplified}.
    \end{split}
    \end{equation} 
Without loss of generality, the adversary can choose $\pm\rho$ to obtain the highest $d$. 
Moreover, using \ref{eq::dsimplified}, we can compute an upper bound for $d$ employing the sensitivity $\Delta_2(q)$ and the Cauchy-Schwarz inequality:  
    \begin{align}
       d 
    &\leq \frac{\Delta_2(q)^2}{\sigma^2(1-\rho^2)}
    + \frac{2|\rho|}{\sigma^2(1-\rho^2)} ||\Re(q(\D)-q(\D'))||_2 \cdot ||\Im(q(\D)-q(\D'))||_2   \\
    &\leq \frac{\Delta_2(q)^2}{\sigma^2(1-\rho^2)} + \frac{2|\rho|}{\sigma^2(1-\rho^2)} \Delta_2(q_{\Re})  \Delta_2(q_{\Im}) := \hat{d}. \label{eq::dtotal} 
    \end{align}
No tighter bound than \ref{eq::dtotal} can be computed without making assumptions on the query $q$ or the databases $\D, \D'$.   
Thus, we can use $\hat{d}$ to compute the trade-off that bounds the worst-case scenario of the cGM:
    \begin{equation}
        f_{\sqrt{\hat{d}}}(\alpha) = \Phi \left(\Phi^{-1}(1-\alpha) -\sqrt{\hat{d}} \right), \label{eq::tradeoffcgm}      
    \end{equation}
where $\alpha$ is the Type-I statistical error and $\Phi$ is the cumulative distribution of the standard, real-valued normal distribution.
To conclude, we note that \ref{eq::tradeoffcgm} is the trade-off function of a (real-valued) $\sqrt{\hat{d}}$-GDP mechanism. 
\end{proof}

\end{document}